\documentclass[11pt,a4paper]{article}
\usepackage{jcappub}
\usepackage{natbib}

\newcommand{\veps}{\varepsilon}
\newcommand{\pat}{\partial}

\title{To theory of asymptotically stable accelerating Universe
in Riemann-Cartan spacetime}
\author[a,b]{A.S. Garkun,}
\author[b]{V.I. Kudin,}
\author[b,c,d]{and A.V. Minkevich}
\affiliation[a]{The National Academy of Sciences of Belarus, Minsk, Belarus}
\affiliation[b]{Belarusian State University, Minsk, Belarus}
\affiliation[c]{Warmia and Mazury University in  Olsztyn, Poland}
\affiliation[d]{Kazan Federal University, Russia}

\emailAdd{garkun@bsu.by} \emailAdd{minkav@bsu.by}
\emailAdd{awm@matman.uwm.edu.pl} \emailAdd{kudzin\_w@tut.by}

\abstract{Homogeneous isotropic cosmological models built in the framework of
the Poincar\'e gauge theory of gravity based on general expression of
gravitational Lagrangian with indefinite parameters are analyzed. Special
points of cosmological solutions for flat cosmological models at asymptotics
and conditions of their stability in dependence of indefinite parameters are
found. Procedure of numerical integration of the system of gravitational
equations at asymptotics is considered. Numerical solution for accelerating
Universe without dark energy is obtained.}


\begin{document}

\maketitle

\section{Introduction}

One of the most principal achievements of observational cosmology is the
discovery of the acceleration of cosmological expansion at present epoch. In
order to explain accelerating cosmological expansion  in the framework  of
General Relativity Theory (GR), the notion of dark energy (or quintessence) as
some hypothetical kind of gravitating matter with negative pressure was
introduced. Then the explanation of cosmological acceleration in the frame of
GR leads to conclusion that approximately 70\% of energy in our Universe is
related to dark energy.

In the frame of standard $\Lambda CDM$-model the dark energy is associated with
cosmological constant $\Lambda$, which is related to the vacuum energy density
of matter fields. In terms of quantum field theory the vacuum energy density
diverges and can be eliminated by means of renormalization procedure. At the
same time the value of cosmological constant $\Lambda$, which is introduced
into gravitational equations of GR manually, is very small and close to average
energy density in the Universe at present epoch.

Another situation takes place in the framework of gravitation theory in the
Riemann-Cartan spacetime $U_4$ - Poincar\'e gauge theory of gravity (PGTG) (see
\cite{mgkJCAP} and Refs herein). At first it should be noted that the PGTG is
natural and in certain sense necessary generalization of metric gravitation
theory by applying the local gauge invariance principle to gravitational
interaction, if the Lorentz group is included into the gauge group which
corresponds to gravitational interaction
\cite{kibble,brodskii,sciama,hehl1,hayashi}. In the frame of PGTG the effective
cosmological constant appears in cosmological equations by virtue of the most
complicated structure of physical spacetime, notably by spacetime torsion
\cite{a12,a19}. As it was shown in \cite{a19}, the physical spacetime in the
vacuum (in absence of gravitating matter) in the frame of PGTG in general case
has the structure of Riemann-Cartan continuum with de Sitter metrics, but not
Minkowski spacetime. Corresponding results were obtained by analyzing isotropic
cosmology built in the frame of PGTG based on general expression of
gravitational Lagrangian ${\cal L}_{\mathrm{g}}$ including both a scalar
curvature and invariants quadratic in the curvature and torsion tensors with
indefinite parameters (see \cite{a19,mgkJCAP} and Refs herein).
\footnote{Similar results were discussed later in \cite{chee1,chee2,qi} by
using the gravitational Lagrangian simplified in comparison with
\cite{a12,a19}.} From the point of view of PGTG the effect of gravitational
repulsion leading to accelerating cosmological expansion at present epoch has
the vacuum origin and it is connected with the change of gravitational
interaction provoked by spacetime torsion without any dark energy.

The principal change of gravitational interaction takes place also in the
beginning of cosmological expansion, when the energy density $\rho$ and
pressure $p$ have extremely high values: by virtue of existence of limiting
energy density, close to which the gravitational interaction in the case of
usual matter satisfying standard energy conditions is repulsive, isotropic
cosmology is regular \cite{a21}. The regularity takes place not only with
respect to energy density and metric characteristics (scale factor of
Robertson-Walker metric, Hubble parameter with its time derivative), but also
with respect to torsion and curvature functions \cite{a22}.  It should be noted
that indicated physical results were obtained by certain restrictions on
indefinite parameters of gravitational Lagrangian ${\cal L}_{\mathrm{g}}$ (see
below). Additional restrictions on indefinite parameters can be found by
investigation presented below of cosmological models for accelerating Universe.

The present paper is devoted to analysis of homogeneous isotropic models (HIM)
with two torsion functions with the purpose to obtain asymptotically stable
solutions for accelerating Universe. At first in Section 2 the principal
relations of isotropic cosmology built in the frame of PGTG and using in this
paper are given.

\section{\label{secii}Principal relations of isotropic cosmology in Riemann-Cartan spacetime}
In the framework of PGTG the role of gravitational field variables play the
tetrad $h^i{}_\mu$ and the Lorentz connection $A^{ik}{}_\mu$; corresponding
field strengths are the torsion tensor $S^i{}_{\mu\nu}$ and the curvature
tensor $F^{ik}{}_{\mu\nu}$ defined as
\[
S^i{}_{\mu \,\nu }  = \partial _{[\nu } \,h^i{}_{\mu ]}  - h_{k[\mu }
A^{ik}{}_{\nu ]}\,,
\]
\[
F^{ik}{}_{\mu\nu }  = 2\partial _{[\mu } A^{ik}{}_{\nu ]}  + 2A^{il}{}_{[\mu }
A^k{}_{|l\,|\nu ]}\,,
\]
where holonomic and anholonomic space-time coordinates are denoted by means of
greek and latin indices respectively.

We will consider the PGTG based on gravitational Lagrangian given in the
following general form
\begin{eqnarray}\label{lagr}%\fl
{\cal L}_{\mathrm{g}}=  f_0\,
F+F^{\alpha\beta\mu\nu}\left(f_1\:F_{\alpha\beta\mu\nu}+f_2\:
F_{\alpha\mu\beta\nu}+f_3\:F_{\mu\nu\alpha\beta}\right)+
F^{\mu\nu}\left(f_4\:F_{\mu\nu}+f_5\:
F_{\nu\mu}\right) \nonumber \\
+f_6\:F^2+S^{\alpha\mu\nu}\left(a_1\:S_{\alpha\mu\nu}+a_2\:
S_{\nu\mu\alpha}\right)
+a_3\:S^\alpha{}_{\mu\alpha}S_\beta{}^{\mu\beta}, %\nonumber
\end{eqnarray}
where $F_{\mu\nu}=F^{\alpha}{}_{\mu\alpha\nu}$, $F=F^\mu{}_\mu$, $f_i$
($i=1,2,\ldots,6$), $a_k$ ($k=1,2,3$) are indefinite parameters, $f_0=(16\pi
G)^{-1}$, $G$ is Newton's gravitational constant (the velocity of light in the
vacuum is equal to 1). Gravitational equations of PGTG obtained from the action
integral $ I = \int {\left({\cal L}_{\mathrm{g}} + {\cal L}_{\mathrm{m}}
\right) \,} h\,d^4 x$, where $h=\det{\left(h^i{}_\mu\right)}$ and ${\cal
L}_{\mathrm{m}}$ is the Lagrangian of gravitating matter, contain the system of
16+24 equations corresponding to gravitational variables $h^i{}_\mu$ and
$A^{ik}{}_\mu$. By using minimal coupling of gravitational field with matter
the sources of gravitational field in PGTG are the energy-momentum and spin
momentum tensors.

In the framework of PGTG the dynamics of any HIM is described by means of three
functions of time $t$: the scale factor of Robertson-Walker metrics $R$ and two
torsion functions $S_{1}$ and $S_{2}$ determining the curvature tensor. The
system of gravitational equations of PGTG for HIM in considered case is reduced
to 4 equations, which allow to obtain the generalization of Friedmann
cosmological equations and equations for torsion functions \cite{a19,mgkJCAP}.
In general case these equations contain five indefinite parameters -- two
parameters connected with terms of ${\cal L}_{\mathrm{g}}$ quadratic in the
torsion tensor ($ a = 2a_1  + a_2  + 3a_3$, $b = a_2  - a_1$) and three
combinations of parameters $f_i$:
\begin{eqnarray}
    f = f_1  + \frac{{f_2 }} {2} + f_3  + f_4  + f_5  + 3f_{6}\, ,
%\hfill
\nonumber\\
  q_1  = f_2  - 2f_3  + f_4  + f_5  + 6f_{6}, \qquad q_2  = 2f_1  - f_2 .
%\hfill \\
\end{eqnarray}
Indefinite parameters have to obey some restrictions under physical and
mathematical reasons. In accordance with \cite{hayashi} gravitational equations
of PGTG based on gravitational Lagrangian (1) satisfy the correspondence
principle with GR and lead in linear approximation in metric and torsion
functions to Einstein gravitational equations, if the following conditions are
satisfied: $a=0$, $4(f_1+\frac{f_2}{2}+f_3)+f_4+f_5=0$ and $\alpha T \ll 1$,
where $\alpha =\frac {f}{3f_0^2}$ ($f>0$) and $T$ is the trace of canonical
energy-momentum tensor \cite{a23,a24}. The first two conditions are necessary
to exclude higher derivatives of metrics from gravitational equations and the
third condition in the form of inequality is valid for usual gravitating
systems, if the parameter $\alpha$ having inverse dimension of energy density
corresponds to extremely high energy densities. \footnote{In the case of HIM
with the only torsion function the value of $\alpha^{-1}$ determines the
limiting energy density in the beginning of cosmological expansion at a bounce
\cite{minkPL80,a9}.} In the frame of isotropic cosmology the condition $a=0$
was used previously in order to exclude higher derivatives of the scale factor
$R$ from cosmological equations. \footnote{It should be noted that isotropic
cosmology with $a \neq 0$ possesses some principal problems \cite{a20}.} Then
cosmological equations and equations for torsion functions contain four
parameters: $b$ and parameters (2.2), which appear in the following
combinations: $\alpha$, $\varepsilon =\frac{q_2}{f}$ and $\omega= \frac {2f -
q_1 - q_2} {f}$. The investigation of physical and mathematical consequences of
isotropic cosmology allows to obtain some restrictions on these parameters. If
the value of $\alpha^{-1}$ corresponds to the scale of extremely high energy
densities, the explanation of accelerating cosmological expansion at present
epoch together with the effect of existence of limiting energy density lead to
the following conditions \cite{a23,a24}: $|\varepsilon|\ll 1$,
$0<1-\frac{b}{f_0}\ll 1$, $0<\omega<4$.

For further analysis, we transform cosmological equations (Eqs (3.1)-(3.2) in
Ref.~\cite{mgkJCAP}) to dimensionless form by introducing dimensionless units
for all variables and parameter $b$ entering these equations and denoted by
means of tilde:
\begin{equation}\label{dimless}
\begin{array}{lcl}
    t\to\tilde{t}=t/\sqrt{6 f_0 \alpha},& {}
            & R\to\tilde{R}=R/\sqrt{6f_0 \alpha},\\
    \rho\to\tilde{\rho}=\alpha\,\rho, & & p\to\tilde{p}=\alpha\,p,\\
    S_{1,2}\to\tilde{S}_{1,2}=S_{1,2}\sqrt{6f_0 \alpha}, & &
            b\to\tilde{b} = b/f_0, \\
    H\to\tilde{H}=H\sqrt{6f_0 \alpha}, & &
\end{array}
\end{equation}
where dimensionless Hubble parameter $\tilde{H}$ is defined  by usual way
$\tilde{H}=\tilde{R}^{-1}\frac{d \tilde{R}}{d \tilde{t}}$. As result
cosmological equations take the following dimensionless form, where the
differentiation with respect to dimensionless time $\tilde{t}$ is denoted by
means of the prime and the sign of \ $\tilde{}$\, is omitted below:
\begin{eqnarray}\label{gcfe1}
    \frac{k}{R^2}  +  (H-2S_1)^2 %\nonumber\\
    &=& \frac{1} {Z}
        \left[
            {\rho  +\left(Z- b\right) S_2^2
            + \frac{1}{4} \left( {\rho  - 3p - 2bS_2^2 } \right)^2 }
        \right]
\nonumber\\
        &&- \frac{\varepsilon}{2Z}
            \left[
                {\left( {HS_2  + S_2' } \right)^2
                + 4\left( {\frac{k}{{R^2 }} - S_2^2 } \right)S_2^2 }
            \right],
\end{eqnarray}
\begin{eqnarray}\label{gcfe2}
    H'  &+&  H^2 - 2HS_1 - 2S_1' %\nonumber \\
     =  -\frac{1} {2Z}
        \left[
            \rho  + 3p - \frac{1 } {2} \left( {\rho  - 3p - 2bS_2^2 } \right)^2
        \right]
\nonumber\\
        &  - & \frac{\varepsilon }{Z}\left( {\rho  - 3p - 2bS_2^2 } \right)S_2^2
%\nonumber\\
         + \frac{{\varepsilon }} {2Z}
            \left[ {\left( {HS_2  + S_2' } \right)^2
                + 4\left( {\frac{k}{{R^2 }} - S_2^2 } \right)S_2^2 }
            \right], %\\
\end{eqnarray}
\[
    \left(Z \equiv 1+\rho - 3p - 2\left( {b + \varepsilon } \right)S_2^2\right).\nonumber
\]
The torsion function $S_{1}$ (Eq. (3.3) in Ref.~\cite{mgkJCAP}) in
dimensionless form in equations (\ref{gcfe1})--(\ref{gcfe2}) is
\begin{equation} \label{S1expr}
    S_1  =   -\frac{3}{4Z} \left\{
           H\left[\left(\rho+p\right)%
        \left(3\frac{dp}{d\rho}-1 \right) + 2(\varepsilon + \frac{\omega}{3}) S_2^2\right]
            -\frac{2}{3}\left( 2b - (\veps + \omega)  \right) S_2 \, S_2'
        \right\}
\end{equation}
and  dimensionless torsion function $S_{2}$ (Eq. (3.4) in Ref.~\cite{mgkJCAP})
satisfies the following differential equation of the second order:
\begin{eqnarray}\label{gcfe3} & &
    \varepsilon \left[ S_2''  + 3H S_2'  + 3H' S_2  - 4\left(S_1'  - 3 HS_1
        + 4S_1^2\right) S_2  \right]
        - 2\left(1-\frac{\omega}{2}\right) \left( {\rho  - 3p - 2bS_2^2 } \right)S_2
        \nonumber \\
& &
        - 2\left( {1  - b}\right)S_2 - 2\omega \left[\frac{k}{R^2}  +  (H-2S_1)^2 - S_2^2 \right]S_2  = 0\,.
\end{eqnarray}
The conservation law for gravitating matter in dimensionless units has the
usual form
%\begin{equation}
\begin{equation}
\label{conslaw} \rho'+3H\left(\rho+p\right)=0.
\end{equation}
%\end{equation}

\section{\label{seciii}Critical points analysis}

The system of equations (\ref{gcfe2}) -- (\ref{gcfe3}) together with
conservation law (\ref{conslaw}) completely determine the dynamics of HIM, if
the equation of state of matter is given. The composition of gravitating matter
and its equation of state change by cosmological evolution. By analysis of HIM
at asymptotics we will consider further flat model ($k=0$) filled with matter
with barotropic equation of state $p=w\rho$ $(w=const)$. The aforementioned
system of equations can be represented in the form of four first order
differential equations for $H$, $S_2$, $U=S_2'$ and $\rho$:
\begin{equation}\label{firstordersyst}
 M_0 \mathbf{Y}'=\mathbf{F},
\end{equation}
where the matrix $M_0$ is
\begin{equation}
 M_0=\left(
    \begin{array}{cccc}
     1-2\frac{\pat S_1}{\pat H} & -2\frac{\pat S_1}{\pat S_2} & -2\frac{\pat S_1}{\pat U} & -2\frac{\pat S_1}{\pat\rho}
        \\
     0 & 1 & 0 & 0
    \\
     3 \varepsilon S_2 - 4\varepsilon\frac{\pat S_1}{\pat H}S_2 & - 4\varepsilon\frac{\pat S_1}{\pat S_2}S_2
     & \varepsilon (1- 4\frac{\pat S_1}{\pat U}S_2)
     & - 4\varepsilon\frac{\pat S_1}{\pat\rho}S_2
    \\
     0 & 0 & 0 & 1
    \end{array}
      \right)
\end{equation}
and
\[
 \mathbf{Y}=\left(
    \begin{array}{l}
     H\\
     S_2\\
     U\\
     \rho
    \end{array}
 \right), \qquad
 \mathbf{F}=\left(
    \begin{array}{l}
     \mathcal{F}_1\\
     \mathcal{F}_2\\
     \mathcal{F}_3\\
     \mathcal{F}_4
    \end{array}
 \right),
\]
where $\mathcal{F}_i$ ($i=1,2,3,4$) are the following functions of
$H,S_2,U,\rho$:
\begin{eqnarray}
  \mathcal{F}_1 & = & -H^2 + 2HS_1
      -\frac{1} {2Z}
        \left\{
            \left(1+3w\right)\rho  - \frac{1 } {2} \left[ \left(1-3w\right)\rho  - 2bS_2^2 \right]^2
        \right\}
\nonumber\\
    & & -  \frac{\varepsilon }{Z}\left[ \left(1-3w\right)\rho  - 2\left(b-1\right)S_2^2  \right]S_2^2
%\nonumber\\
        %& &
        + \frac{{\varepsilon }} {2Z}
            \left( {HS_2  + U } \right)^2,\\
  \mathcal{F}_2 & = & U,\\
  \mathcal{F}_3 & = & - \varepsilon \left[3H U  + 4\left(3 H
        - 4S_1\right) S_1 S_2  \right]
        + 2\left(1-\frac{\omega}{2}\right)\left[\left(1-3w\right) \rho  - 2bS_2^2  \right]S_2
\nonumber \\
    & &
        + 2\left( 1  - b\right)S_2
        + 2\omega \left[(H-2S_1)^2 - S_2^2\right]S_2 ,\\
  \mathcal{F}_4 & = & -3 \left(1+w\right)\rho H
\end{eqnarray}
and the function $S_1$ takes the form as
\begin{equation} \label{S1expr2}
    S_1  =   -\frac{3}{4Z} \left\{
           H\left[\left(1+w\right)%
        \left(3w-1 \right)\rho + 2(\varepsilon + \frac {\omega}{3})S_2^2\right]
            -\frac{2}{3}\left( {2b - (\veps + \omega) } \right) S_2 U
        \right\},
\end{equation}
\[
    \left(Z = 1+\rho (1- 3w) - 2\left( {b + \varepsilon } \right)S_2^2\right).\nonumber
\]

Critical points $P_i=P_i(H_\mathrm{c}, S_{2\mathrm{c}}, U_\mathrm{c},
\rho_\mathrm{c})$ of the first order system of differential equations
(\ref{firstordersyst}) can be obtained by setting $H'$, $S_2'$, $S_1'$, $\rho'$
to zero \cite{agarwal,arnold}, i.e. by solving the following system of
equations:
\begin{equation}\label{specpoint}
 \mathcal{F}_i(H, S_2, U, \rho)=0 \qquad \left(i=1,\ldots,4\right).
\end{equation}
From (3.4) follows that $U_\mathrm{c}=0$. In the case of considering flat model
solutions of (\ref{specpoint}) have to satisfy (\ref{gcfe1}) with $k=0$.

Obviously, the point  $P_0$ with vanishing values of $H_\mathrm{c},
S_{2\mathrm{c}}, \rho_\mathrm{c}$ satisfies (\ref{specpoint}). Appropriate
solution with vanishing $S_2$-function at asymptotics appears at specific
choice of parameters and does not have physical interest \cite{a19}.
Analogously to GR this point is the point of complicated equilibrium. To
analyze the stability of other critical points
$P(H_\mathrm{c},S_{2\mathrm{c}},0,\rho_\mathrm{c})$ satisfying
(\ref{specpoint}) it is necessary to build linearized form of the system
(\ref{firstordersyst}). Near the critical point the variables can be written in
the form $H=H_\mathrm{c}+\Delta H$, $S_2=S_{2\mathrm{c}}+\Delta S_2$, $U=\Delta
U$, $\rho=\rho_\mathrm{c}+\Delta\rho$ and the linearization of the system
(\ref{firstordersyst}) takes the following relation
\begin{equation}\label{firstordersystapprox}
 \Delta\mathbf{Y}'=M_0^{-1}M_1\,\Delta\mathbf{Y},
\end{equation}
where the matrix $M_0^{-1}$ is taken on the point $P$ and the components of the
matrix $M_1$ are given by
\[
 M_{1,ij}=\left.\left(\frac{\pat \mathcal{F}_i}{\pat Y_j}\right)\right|_{P}.
\]
Stability of the point $P$ is determined by the eigenvalues $\lambda_i$ of the
matrix $M_0^{-1}M_1$ \cite{agarwal,arnold}. Characteristic equation
$\det\left(M_1-\lambda M_0\right)=0$ leads to quartic expression with respect
to $\lambda$, which can be written as
\begin{equation}\label{charactereq}
  \lambda^4+c_1\lambda^3+c_2\lambda^2+c_3\lambda+c_4=0,
\end{equation}
where $c_i$ ($i=1,2,3,4$) are some functions of indefinite parameters. If the
real parts of all $\lambda_i$ are negative, then the critical point $P$ is
stable and the gravitational equations (\ref{gcfe2}) -- (\ref{conslaw}) have
solution with asymptotics to this point $H\to H_\mathrm{c}$, $S_2\to
S_{2\mathrm{c}}$, $S'_2 \to 0$, $\rho \to \rho_\mathrm{c}$ at $t\to +\infty$.

According to the Routh-Hurwitz theorem  all $\lambda_i$ will have negative real
parts if the main minors of the matrix
\begin{equation}\label{RHmatrix}
 \left(
  \begin{array}{cccc}
   c_1 & 1 & 0 & 0\\
   c_3 & c_2 & c_1 & 1\\
   0 & c_4 & c_3 & c_2 \\
   0 & 0 & 0 & c_4
  \end{array}
 \right)
\end{equation}
are positive \cite{agarwal}, i.e.
\begin{equation}\label{stabilityineq}
 c_1>0,\qquad c_1c_2-c_3>0,\qquad c_1c_2c_3-c_1^2c_4-c_3^2>0\qquad \text{and}\qquad c_4>0.
\end{equation}

The equation $\mathcal{F}_4(H_\mathrm{c},S_{2\mathrm{c}}, 0,\rho_\mathrm{c})=0$
is satisfied if at least one of possibilities is fulfilled: $H_\mathrm{c}=0$ or
$\rho_\mathrm{c}=0$. Simultaneous fulfillment of these conditions leads to
trivial solutions with $S_{2\mathrm{c}}=0$. Because the energy density in the
case of considering flat model tends at asymptotics to zero, the physical
interest assume critical points with non-vanishing Hubble parameter and
vanishing energy density.

If $\rho_\mathrm{c}=0$ the system (\ref{specpoint}) is reduced to the system of
two algebraic equations
\begin{eqnarray}
  & & -H^2 + 2HS_1
      +\left[b^2  +  2\varepsilon\left(b-1\right)\right]\frac{S_2^4}{Z}
      + \frac{{\varepsilon }} {2Z} H^2S_2^2=0,
 \label{specpoint2a}\\
  & &    \left[2\varepsilon \left(3 HS_1 - 4S_1^2\right) + 2 b(1-\frac{\omega}{2}) S_2^2
  -\omega ( (H-2S_1)^2-S_2^2) - \left(1  - b\right)\right]S_2 = 0,
  \label{specpoint2b}
\end{eqnarray}
where the functions $S_1$ and $Z$ can be represented in the following form
\begin{equation} \label{S1expr2st1}
    S_1  =   -\frac{3}{2Z}(\varepsilon + \frac{\omega}{3}) H S_2^2
    \qquad \text{and} \qquad
    Z = 1 - 2\left( {b + \varepsilon } \right)S_2^2.
\end{equation}
Neglecting the case $S_2=0$ and by using (\ref{S1expr2st1}) the system of
equations (\ref{specpoint2a})--(\ref{specpoint2b}) can be rewritten in the
following form
\begin{eqnarray}
 & &
 H^2=\frac{b^2-2\veps\left(1-b\right)}{1+\left(\frac{1}{2}\veps+\omega-2b\right)S_2^2} S_2^4
 \qquad
  \left( 1+\left(\frac{1}{2}\veps+\omega-2b\right)S_2^2\neq 0\right),
 \label{specpoint2a1}\\
  & &
  9\varepsilon (\varepsilon+\frac{\omega}{3})\left[1-2(b-\frac {\omega}{3})S_2^2\right]H^2S_2^2
  +\omega H^2 [1+(\varepsilon -2b +\omega) S_2^2]^2
  \nonumber \\
  & &
    -[1 - 2\left( {b + \varepsilon } \right)S_2^2]^2 \left[S_2^2(\omega (1-b)+2b)-(1-b)\right] = 0.
  \label{specpoint2b1}
\end{eqnarray}
Then Eqs.(\ref{specpoint2a1}) -- (\ref{specpoint2b1}) allow to obtain the
equation for $S_2$ in closed form:
\begin{eqnarray}
\label{S2closed}
& &
 \left[ 1-(2 b+\omega+4\varepsilon )S_2^2\right]
\nonumber \\
& & \times
 \left\{2 \left[2 b^3 (\omega -4)
    +b^2 \left(-\omega ^2+4 \omega  (\varepsilon +1)+\varepsilon  (9 \varepsilon +2)\right)
\right. \right. \nonumber \\
& & \left. \left. \phantom{\times}
    -2 b (\omega +2) \varepsilon  (\omega +2 \varepsilon )
    +2 \omega  \varepsilon  (\omega +2 \varepsilon )\right] S_2^6
 + \left(-8 b^3+2 b^2 (\omega +\varepsilon +12)
\right. \right. \nonumber \\
& & \left. \left. \phantom{\times}
    -4 b \left(\omega  (\varepsilon +2)+2 \varepsilon ^2+\varepsilon \right)
    +4 \varepsilon  (\omega +2 \varepsilon )\right) S_2^4
\right. \nonumber \\
& & \left. \phantom{\times}
 + \left(8 b^2-b (2 \omega +\varepsilon +12)+2 \omega +\varepsilon \right) S_2^2
 -2 (b-1)\right\}=0,
\end{eqnarray}
and also the equation for $H$ in closed form
\begin{eqnarray}\label{criticalH}
 & &
 \varepsilon  (\omega +2 \varepsilon ) \left[2 b^3 (\omega -4)+b^2 \left(-\omega ^2+4 \omega  (\varepsilon +1)
 +\varepsilon  (9 \varepsilon +2)\right)
    \right. \nonumber \\
 & & \left. \phantom{\left\{ \right\} }
 -2 b (\omega +2) \varepsilon  (\omega +2 \varepsilon )
 +2 \omega  \varepsilon  (\omega +2 \varepsilon )\right] H^6
 +\left[-2 b^4 \left(\omega ^2+2 \omega  (\varepsilon -2)+2 (\varepsilon -2) \varepsilon \right)
    \right. \nonumber \\
 & & \left. \phantom{\left\{ \right\} }
    +b^3 \left(2 \omega ^3+\omega ^2 (\varepsilon-8)-2 \omega  \varepsilon  (5 \varepsilon +9)
        -\varepsilon ^2 (17 \varepsilon +22)\right)
    \right. \nonumber \\
 & & \left. \phantom{\left\{ \right\} }
    +b^2 \left(\omega ^3 (4 \varepsilon -2)+\omega^2 (\varepsilon  (18 \varepsilon -1)+8)
        +2 \omega  \varepsilon  (\varepsilon  (18 \varepsilon +11)+12)
        +\varepsilon ^2 (\varepsilon  (32 \varepsilon +41)+34)\right)
    \right. \nonumber \\
 & & \left. \phantom{\left\{ \right\} }
    -4 b \varepsilon  (\omega +2 \varepsilon ) \left(2 \omega ^2+5 \omega  \varepsilon
        +\varepsilon  (8 \varepsilon +3)\right)
    +2 \varepsilon  (\omega +2 \varepsilon ) \left(2 \omega ^2+5 \omega  \varepsilon +8 \varepsilon ^2\right)
 \right] H^4
    \nonumber \\
 & & \phantom{\left\{ \right\} }
-2 \left(b^2+2 (b-1) \varepsilon \right) \left[4 b \left(b^2-b \omega +\omega
\right)+(b-1) \varepsilon  (b (b-4 \omega -2)+4 \omega )-8 (b-1)^2 \varepsilon
^2\right] H^2
    \nonumber \\
 & & \phantom{\left\{ \right\} }
+2 (b-1)^2 \left(b^2+2 (b-1) \varepsilon \right)^2=0.
\end{eqnarray}

\subsection{Approximate analysis in the case $ 0<1-b\ll 1 $}

Analytic analysis of stable points determined by the system
(\ref{specpoint2a1})--(\ref{specpoint2b1}) is possible approximately only if
$1-b\to +0$ and $\varepsilon\to 0$. In other cases it is necessary to use
numerical methods. By supposing that values of dimensionless functions $S_2$
and $H$ at asymptotics in (\ref{specpoint2a1})--(\ref{specpoint2b1}) are small
($|S_2|\ll 1$, $|H|\ll 1$) it is easy to obtain the following approximate
solution of equations (\ref{specpoint2a1}) -- (\ref{specpoint2b1}) if $0<1-b\ll
1$ and $|\varepsilon|\ll 1$:
\begin{equation}
H_\mathrm{c}= \frac{1-b}{2\sqrt{b}}, \qquad
S_{2\mathrm{c}}=\sqrt{\frac{1-b}{2b}}.
\end{equation}
 This solution was obtained initially in ref.~\cite{a12}.
The stability of the critical point $P_1\approx\left(\frac{1-b}{2\sqrt{b}},
\sqrt{\frac{1-b}{2b}},0,0\right)$ can be analyzed analytically. The matrix
 $M_0$ and $M_1$ in this case according to their definition have the following form:
\[
 M_0=
 \left(
 \begin{array}{ccccccc}
  1+\frac{1}{2}(1-b) \left(\omega+3 \varepsilon \right) & \quad &
    0 & \quad & \sqrt{\frac{1-b}{2}} (\omega +\varepsilon -2) & \quad &
    -\frac{3}{4}(1+w) (1-3w) (1-b) \\
  0 & & 1 && 0 && 0 \\
  3\sqrt{\frac{1-b}{2}}\varepsilon && 0 && \varepsilon +(1-b) \left(\varepsilon ^2+\omega  \varepsilon -2 \varepsilon \right) && 0 \\
  0 && 0 && 0 && 1
 \end{array}
\right),
\]
\[
 M_1=
 \left(
  \begin{array}{ccccccc}
    -(1-b) & \quad & 0 & \quad & 0  & \quad & -\frac{3}{2}w-(1-b) (\varepsilon +1)-\frac{1}{2} \\
    0 && 0 && 1 && 0 \\
    0 && -4 (1-b) && -\frac{3}{2} (1-b) \varepsilon  && -\sqrt{\frac{1-b}{2}} (\omega -2) (1-3 w) \\
    0 && 0 && 0 && -\frac{3}{2} (1-b) (w +1)
  \end{array}
\right)
\]

As result we obtain the characteristic polynomial (\ref{charactereq}) in the form
\begin{eqnarray}
%& &
    \varepsilon\left[\lambda+\frac{3}{2}(1+w)(1-b)\right]
      \left[\lambda^3+\frac{5}{2}(1-b)\lambda^2+4\frac{1-b}{\varepsilon}\lambda+4\frac{(1-b)^2}{\varepsilon}\right]=0,
\end{eqnarray}
where higher order terms in powers of $(1-b)$ are omitted. Due to factorization
of this equation the analysis of real parts of $\lambda$ is reduced to the
analysis of cubic equation $\lambda^3+c_1\lambda^2+c_2\lambda+c_3=0$. The
Routh-Hurwitz theorem  in this case requires: $c_1>0$, $c_1c_2-c_3>0$ and
$c_3>0$. As result we obtain:
\begin{equation}
 \varepsilon>0, \quad w>-1.
\end{equation}

\subsection{Numerical analysis of stability}

\begin{figure}[t]
\begin{minipage}{0.48\textwidth}\centering{
 \includegraphics[width=\linewidth]{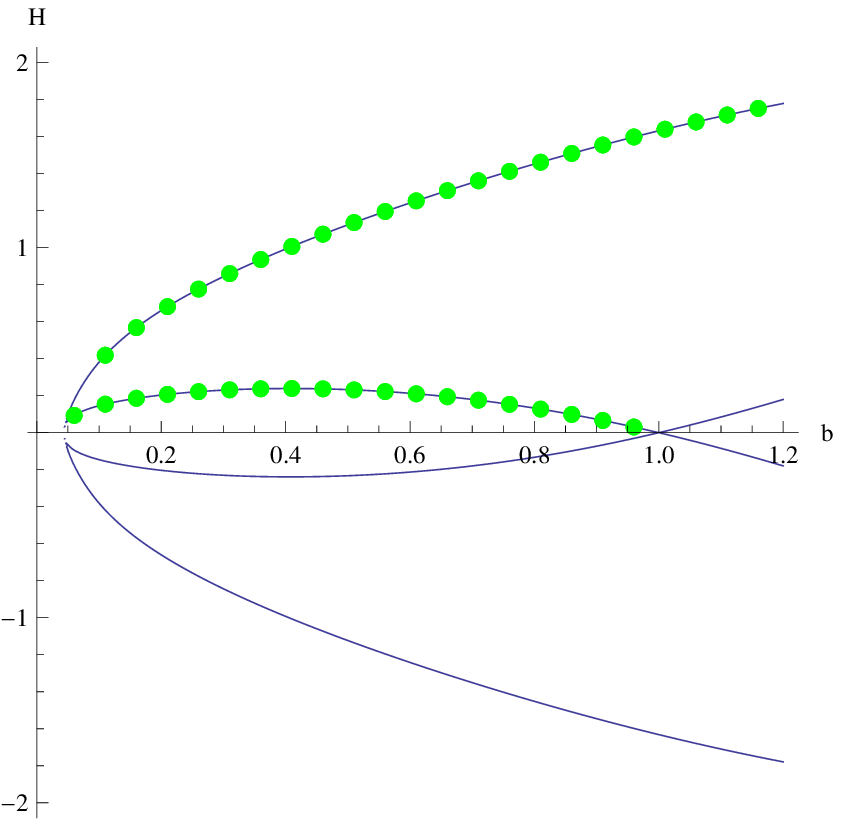}
}
\end{minipage}\, \hfill\,
\begin{minipage}{0.48\textwidth}\centering{
 \includegraphics[width=\linewidth]{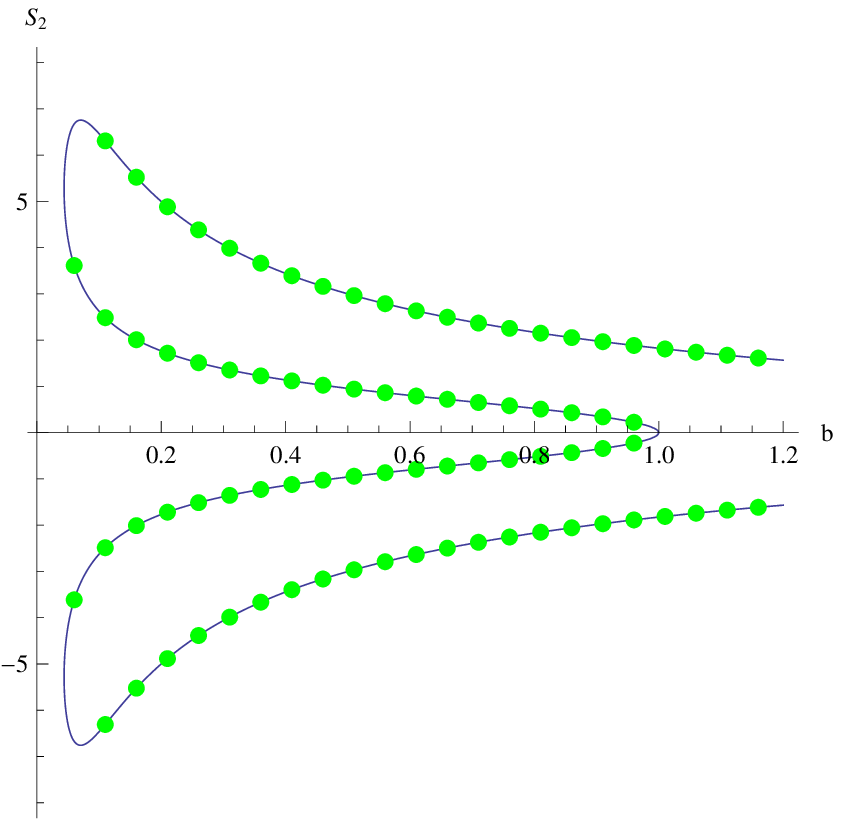}
}
\end{minipage}
\caption{\label{figstable1}$H_\mathrm{c}$ and $S_{2\mathrm{c}}$ as function of
$b$ ($\varepsilon=0.001$, $\omega=2.5$). Stable critical points are marked by
(green) circle. Unstable critical points and points of complicated equilibrium
are not shown in this figure. Solid lines in the left and right panel of figure
correspond to solutions of eqs. (\ref{S2closed}) and (\ref{criticalH}).}
\end{figure}

As an exact analytic expression for solution of the system
(\ref{specpoint2a1})--(\ref{specpoint2b1}) does not exist in general case, it
is necessary to use numerical methods to analyze stability of the critical
points. The procedure of the numerical analysis of the stability is following.
\begin{enumerate}
 \item For given values of $\varepsilon$ and $\omega$, the system (\ref{specpoint2a1})--(\ref{specpoint2b1})
 is solved numerically for the set of values $b$.
 \item For every real solution of the system (\ref{specpoint2a1})--(\ref{specpoint2b1}) at given values
 of $\varepsilon$, $\omega$ and $b$ characteristic equation $\det\left(M_1-\lambda M_0\right)=0$ has to be solved
 with respect to $\lambda$.
 \item The real parts of obtained $\lambda_i$ have to be tested for negativity.
\end{enumerate}
For example, the results of this procedure for $\varepsilon=0.001$ and
$\omega=2.5$ are given in figure~\ref{figstable1}. The calculation are
performed for $b$ varying from $0{.}01$ to $1{.}2$ with a step $\Delta b=0.05$.
 In the right panel of figure~\ref{figstable1} the curves for $S_2$ determined by (\ref{S2closed}) are imposed. In the
left panel of this figure the curves for $H$ determined by (\ref{criticalH})
are imposed.

From figure~\ref{figstable1} one can see, that there is minimal value of $b$
assuming nontrivial solution of eqs.
(\ref{specpoint2a1})--(\ref{specpoint2b1}). As numerical analysis shows, this
value depends on parameter $\varepsilon$ and weakly depends on $\omega$. It
follows  from (\ref{specpoint2a1}) that for sufficiently small $\veps$ and
positive values of $\omega$ we have $b^2-2\veps(1-b)>0$. As result  we obtain
the following restriction on $b$
\begin{equation}
 b>-\varepsilon+\sqrt{\varepsilon\left(2+\varepsilon\right)}.
\end{equation}

Among various cosmological solutions of PGTG with stable asymptotics there are
solutions which can correspond to observable Universe at present epoch. Such
solutions we obtain by using the following restrictions on indefinite
parameters: $0<\varepsilon\ll 1$, $0<1-b\ll 1$ and the parameter $\omega$ has
to satisfy the condition $0< \omega <4$ \cite{a23,a24}. It should be noted that
there are two different solutions at such restrictions on parameters, and only
one of these solutions with small values of $S_2^2$ and $H$ is physically
acceptable \cite{a21}.

\section{Late-time approximation of cosmological solution}\label{nintsec}

Now we will analyze the late-time behaviour of the solution of the system
(\ref{gcfe2})--(\ref{conslaw}). To make comparison with $\Lambda CDM$-model of
GR we will perform numerical integration of the system of the gravitational
equations for dust matter ($w=0$). To simulate late-time behaviour the initial
conditions will be taken at the point $t_0=0$, which belongs to epoch of
accelerating cosmological expansion. The total procedure includes the following
steps.
\begin{enumerate}
 \item For given acceptable values of $\varepsilon$, $\omega$ and $b$
algebraic system (\ref{specpoint2a1})--(\ref{specpoint2b1}) is solved
numerically and all critical points
 $P_i(H_\mathrm{c}, S_{2\mathrm{c}},0,0)$ are found.  Only real solutions  are considered.
 \item For every critical point, the stability analysis is carried out according to the previous subsection and stable
 point with minimal  positive $H_\mathrm{c}$ and positive (or negative) $S_{2\mathrm{c}}$ is selected.\footnote{These
 values of $H_\mathrm{c}$ and  $S_{2\mathrm{c}}$ correspond to the vacuum as de Sitter spacetime
 with torsion \cite{a19,a21}.}
 \item The torsion function $S_2$ and the Hubble parameter $H$ at late-time approximation can be represented in the form
 \begin{eqnarray}
  \label{Hsubst1}\label{Happrox}
  & & H^2=H_\mathrm{c}^2+y_1\,\rho,\\
  \label{S2subst2}
  & & S_2^2=S_{2\mathrm{c}}^2+y_2\,\rho,
 \end{eqnarray}
 with some coefficients $y_1$ and $y_2$.  As the stable point is selected, then $\rho$ tends to zero at $t\to +\infty$.
 Keeping linear terms in $\rho$  the conservation law (\ref{conslaw}) can be written as
 \begin{equation}\label{conslawapprox}
  \rho'=-3H_{\mathrm{c}}\rho.
 \end{equation}
 Substitution of (\ref{Hsubst1})--(\ref{conslawapprox}) into (\ref{gcfe2})--(\ref{gcfe3}) together with keeping terms linear in $\rho$
 gives two algebraic equations for determination of $y_1$ and $y_2$. Numerical solution of these algebraic
 equations for given  $\varepsilon$, $\omega$, $b$, $H_\mathrm{c}$ and $S_{2\mathrm{c}}$ gives $y_1$ and
 $y_2$.
 \item Positivity of obtained values of $y_1$ and $y_2$ is considered as applicability of the late-time approximation (\ref{Hsubst1})--(\ref{S2subst2})
 and successful choice of stable critical point made in step 2 of current procedure. Further steps are performed only if $y_1>0$ and
 $y_2>0$.
 \item Initial condition for $\rho_0=\rho(t_0)$ is taken from the following equation
 \begin{equation}
  \frac{H^2(t_0)}{H_\mathrm{c}^2}\equiv\frac{H_\mathrm{c}^2+y_1\rho_0}{H_\mathrm{c}^2}=\frac{1}{\Omega_\Lambda},
 \end{equation}
 as result we have $H_0=H(t_0)=\sqrt{H_{\mathrm{c}}^2+y_1\,\rho_0}$ and $S_{20}=S_2(t_0)=\sqrt{S_{2\mathrm{c}}^2+y_2\,\rho_0}$.
 Here $\Omega_\Lambda$ is an additional free parameter that specifies initial conditions.
 \item Initial condition for $S_{20}'=S_2'(t_0)$ is obtained from (\ref{gcfe1}) with $k=0$.
 The minimal in modulus value   of $S_{20}'$ is taken as the initial value.
 \item For this choice of the parameters $\varepsilon$, $\omega$, $b$ and initial conditions $\rho_0$, $H_0$, $S_{20}$ and $S_{20}'$ the system
 of differential equations (\ref{gcfe2})--(\ref{conslaw}) is integrated numerically.
\end{enumerate}

\begin{figure}[t]
\begin{minipage}{0.48\textwidth}\centering{
 \includegraphics[width=\linewidth]{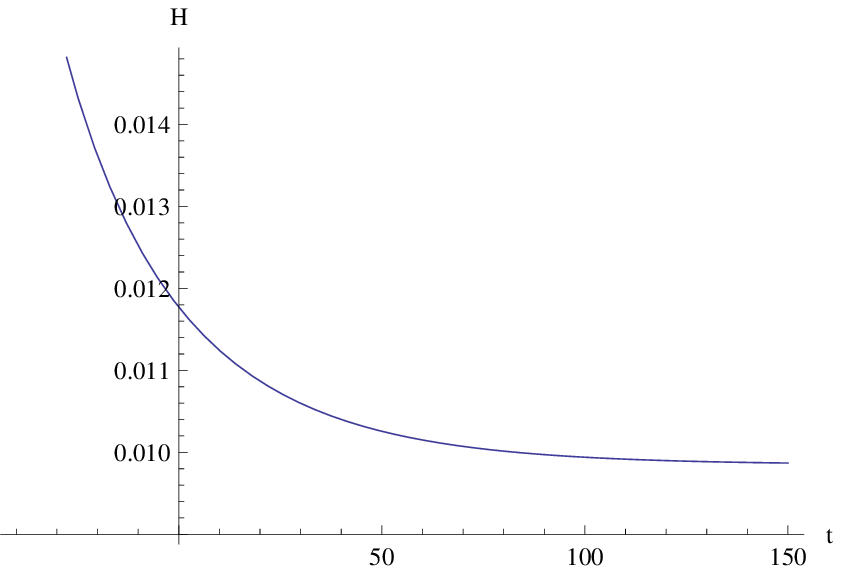}
}
\end{minipage}\, \hfill\,
\begin{minipage}{0.48\textwidth}\centering{
 \includegraphics[width=\linewidth]{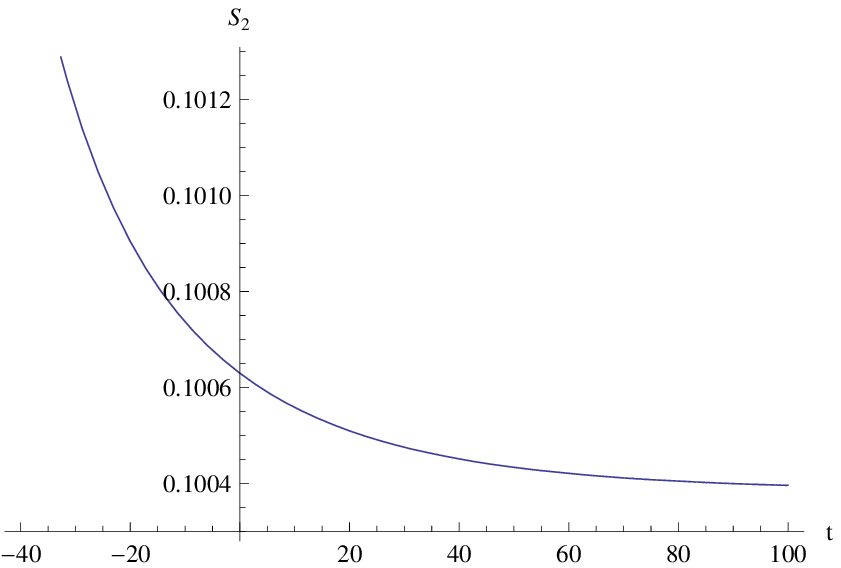}
}
\end{minipage}
\caption{\label{figm1}Late-time behaviour of Hubble parameter and $S_2$ torsion
function.}
\end{figure}

\begin{figure}[t]
\begin{minipage}{0.48\textwidth}\centering{
 \includegraphics[width=\linewidth]{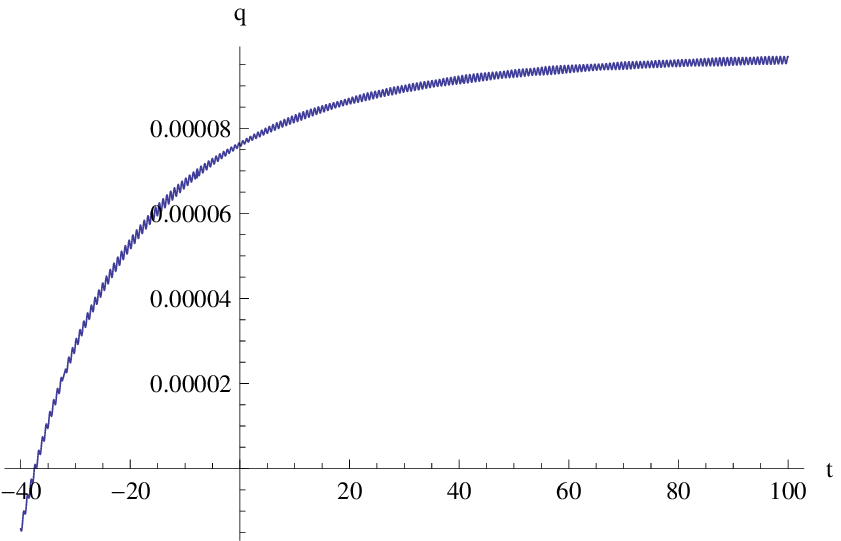}
}
\end{minipage}\, \hfill\,
\begin{minipage}{0.48\textwidth}\centering{
 \includegraphics[width=\linewidth]{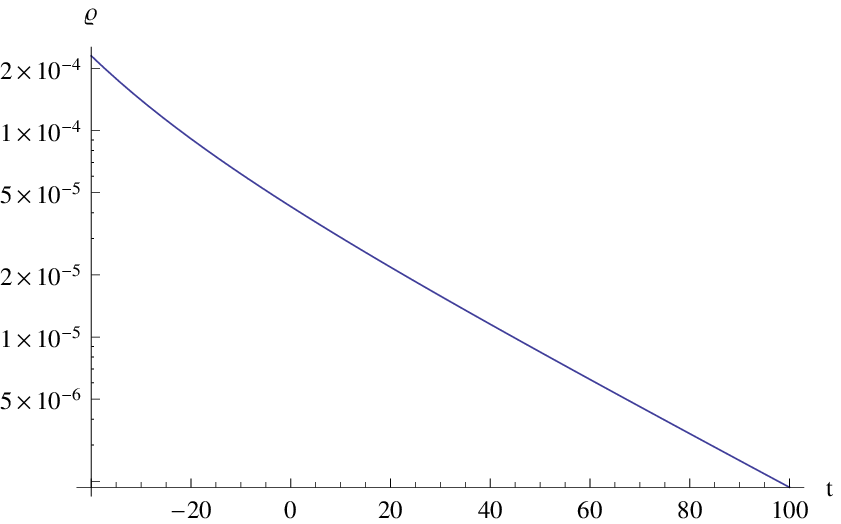}
}
\end{minipage}
\caption{\label{figm2}Late-time behaviour of the acceleration parameter and
energy-density.}
\end{figure}

As an example let us consider the numerical solution at the following
parameters and initial conditions: $\veps=0.001$, $\omega=2.5$, $b=0.98$,
$H_0=0.0118$, $S_{20}=0.1006$, $S'_{20}=-9.5\times 10^{-6}$, $\rho_0=0.000043$.
This choice of the initial conditions gives
$H^2(t_0)/H^2(\infty)=1/\Omega_\Lambda=1/0{.}7$. Figures~2--3 show the
characteristic behaviour of Hubble parameter $H$, torsion function $S_2$,
acceleration parameter $q=R''R/{R'}^2$ and energy density of dust matter $\rho$
for late-time phase of flat cosmological model. As one can see from Figure~3
for acceleration parameter, there was in the past the moment when $q=0$ and the
transition from deceleration to acceleration of cosmological expansion took
place.

Obtained numerical solution for the Hubble parameter and energy density is
close to that of standard $\Lambda CDM$-model. Certain distinction appears in
the behaviour of acceleration parameter $q$ because of its small oscillations
which reduce by decreasing of parameter $\veps$ and disappear if $\veps=0$.

\section{Conclusion}

As follows from our analysis, isotropic cosmology built in the framework of the
Poincar\'e gauge theory of gravity based on general expression of gravitational
Lagrangian leads by certain restrictions on indefinite parameters to
asymptotically stable cosmological solutions for flat homogeneous isotropic
models filled by dust matter, which can describe the stage of accelerated
cosmological expansion of the Universe at present epoch without any dark
energy. The spacetime in asymptotics in obtained solutions has the structure of
Riemann-Cartan continuum with de Sitter metrics and non-vanishing torsion that
demonstrates the dynamical role of the physical vacuum in the frame of PGTG .

\acknowledgments

This work was supported  by a grant from the Belarusian Republican Foundation for Fundamental Research.

\end{document}